\documentclass[12pt]{article}

\def\be{\begin{equation}}
\def\ee{\end{equation}}

\usepackage{euscript,latexsym}

\begin{document}

\title{Bell's Theorem and Random Variables}
\author{Igor Volovich and Yaroslav Volovich\\
~\\
{\it Steklov Mathematical Institute}\\
{\it Gubkin St.8, GSP-1, 117966, Moscow, Russia}\\
{\tt volovich@mi.ras.ru}\\
~\\
{\it Physics Department, Moscow State University}\\
{\it Vorobievi Gori, 119899 Moscow, Russia}\\
{\tt yaroslav\_v@mail.ru}}
\date{~}
\maketitle

\begin{abstract}

Bell's theorem states that quantum correlation function of  two spins can
not be represented as an expectation value of two classical  random
variables. Spin is described in Bell's model by a {\it single scalar} 
random variable. We discuss another classical model of spin in which  spin 
is described by a {\it triple} of classical random variables. It is shown
that in this model the quantum correlation function can be represented as
the expectation value of classical random variables. Implications of this
result to the problem of local causality of quantum mechanics and relations
with problems of moments are briefly mentioned.

\end{abstract}

\newpage
\section{Introduction}

Bell's theorem \cite{Bel} states that there are quantum correlation
functions that can not be represented as an average product of classical
random observables. More specifically, the quantum correlation function of
two spins can not be represented as the expectation value of two classical
{\it scalar} random variables. There are many discussions of Bell's
theorem, see for example \cite{Bell2}-\cite{Hal} for some recent
references.

One often says that Bell deduced his inequality  from realism and locality
\cite{AfrSel}. But in fact Bell uses not only realism and locality but also
a special classical model of spin. Spin is described  in Bell's model by a
{\it single scalar } classical random variable. However it is well known
that spin operator has three components (three Pauli matrices). Therefore
it is more natural to consider another classical model of spin in which
spin is described by means of a  {\it triple} of classical random
variables.

The aim of this note is to show that if one uses the new classical model of
spin then the quantum correlation function of two spins can be represented
as the expectation value of local classical random variables. This should
be contrasted with an  interpretation of Bell's theorem according to which
one can not reproduce the quantum correlation function with classical
probabilistic local model.

The paper is organized as follows. In the next section we remind the
familiar derivation of Bell's theorem. Then we describe the new classical
model of spin and show that one can reproduce the quantum correlation
function of two spins with classical local random variables. Finally,  a
relation of this result with the problem of moments and some its
implications to the problem of local causality of quantum mechanics are
briefly discussed.

\section{Bell's Theorem}

Bell's theorem says that the quantum-mechanical correlation  function of
two spins
\be
\label{eqn5x}
Q(a,b)=\left<\psi|\sigma\cdot a
\otimes\sigma\cdot b|\psi\right>=-a\cdot b
\ee
can not be represented in the form
\be
\label{eqn6x}
Q(a,b)=\int f^{(1)}(a,\omega) f^{(2)}(b,\omega) dP(\omega)
\ee
Here $a=(a_1,a_2,a_3)$ and $b=(b_1,b_2,b_3)$ are unit $3$-vectors, 
$\sigma\cdot a=\sigma_i
a_i$ where $\sigma_i$ are Pauli matrices and
$$
\left|\psi\right>=\frac{1}{\sqrt{2}}
(\left|1,-1\right>-\left|-1,1\right>)
$$
is a singlet state with spin $0$. (We use as spin operators $\sigma_i$
instead of $\sigma_i/2$ just for simplicity of writing. One can make
obvious rescaling to get the standard spin $1/2$ values.) Functions
$f^{(1)}$  and $f^{(2)}$ should satisfy
$$
|f^{(k)}(a,\omega)|\leq 1, ~~k=1,2
$$
and $dP(\omega)$ is a positive
measure on some space $\Omega$ with $\int\! dP(\omega)=1$  (such measure is
called the probability measure). In quantum mechanics  the space $\Omega$
is sometimes called the space of hidden variables. 

In this approach quantum spin is represented by the random variable (field)
$f(a,\omega)$ that takes values $\pm 1$. This description we call Bell's
model.

Let us remind that the familiar proof of Bell's theorem uses the following
inequality \cite{Bell2}
\be
\label{eqn7x}
|C(a,b)-C(a,b')+C(a',b)+C(a',b')|\leq 2
\ee
where
$$
C(a,b)=\int f^{(1)}(a,\omega) f^{(2)}(b,\omega) dP(\omega)
$$
is the classical correlation function.  One can not set
$$
C(a,b)=Q(a,b)=-ab
$$
because there exist such vectors $(ab=a'b=a'b'=-ab'=\sqrt{2}/2)$ for which
one has
\be
\label{eqn8x}
|Q(a,b)-Q(a,b')+Q(a',b)+Q(a',b')|=2\sqrt{2}
\ee
The last equality (\ref{eqn8x}) contradicts to  (\ref{eqn7x}) and this
proves Bell's theorem.

\section{Spin as a Random Variable}

One concludes from Bell's theorem that Bell's model $f(a,\omega)$ of spin
contradicts to quantum mechanics. In this section we describe another
classical model of spin.

An interpretation of relation (\ref{eqn6x}) is that one observer measures
a projection of spin of a particle along vector $a$ while in a distant
region of space a second observer measures the projection of spin of the
second particle along vector $b$. Results of the measurements of the first
observer are represented by a random variable $f^{(1)}(a,\omega)$ and
results of the measurements of the second observer are represented by a
random variable $f^{(2)}(a,\omega)$. The crucial restriction is that
$|f^{(k)}|\leq 1$, $k=1,2$. Actually one can  reduce problem to the case
$f^{(k)}=\pm 1$, $k=1,2$.

It seems to us that it is not natural to describe quantum spin by means of
the classical scalar random function $f(a,\omega)$. There are three
components of the spin operator (Pauli matrices $\sigma_1$, $\sigma_2$ and
$\sigma_3$) and therefore one tends to  describe spin by  using three
classical random variables $\xi_1(\omega)$, $\xi_2(\omega)$ and
$\xi_3(\omega)$ that take values $\pm 1$. Let us show that there are random
variables $\xi^{(1)}_i(\omega)$ and $\xi^{(2)}_i(\omega)$ such that
\be
\label{eqn0x}
\int \xi^{(1)}_i(\omega) \xi^{(2)}_j(\omega) dP(\omega)=-\delta_{ij}
=\left<\psi|\sigma_i \otimes\sigma_j |\psi\right>
\ee

We choose the segment $[0,1]$ as the space $\Omega$ with the measure
$dP(\omega)=d\omega$ and set
$$
\xi_1(\omega)=1,
$$
$$
\xi_2(\omega)=
 \left\{
 \begin{array}{rl}
       1,& \omega\in(\frac{1}{4},
       \frac{1}{2})~or~\omega\in(\frac{3}{4},1)\\
      -1,& otherwise
 \end{array}
 \right.
$$
$$
\xi_3(\omega)=
 \left\{
 \begin{array}{rl}
       1,& \omega\in(0,\frac{1}{2})\\
      -1,& otherwise
 \end{array}
 \right.
$$
Then one has
$$
\int\limits^1_0 \xi_i(\omega)\xi_j(\omega)d\omega=\delta_{ij}
$$
Now if we take
$$
\xi^{(1)}_i(\omega)=\xi_i(\omega), ~~\xi^{(2)}_i(\omega)=-\xi_i(\omega)
$$
then we obtain (\ref{eqn0x}).

Having random variables $\xi^{(1)}_i$ and $\xi^{(2)}_i$ one  can represent
the quantum correlation function as the expectation value of classical
random variables
\be
\label{eq00x}
\left<\psi|\sigma_i a_i\otimes\sigma_j b_j|\psi\right>=-a_i b_i=
\int\limits^1_0 \xi^{(1)}_i(\omega)a_i\xi^{(2)}_j(\omega)b_jd\omega
\ee
Let us stress that we do not interpret
$\xi_i(\omega) a_i$ as describing results of individual measurements
along  an arbitrary vector $a$. In this model individual measurements
are described by random variables $\xi_i(\omega)$ which correspond
to a fixed system of coordinates. One uses  $\xi_i(\omega) a_i$
only to compute the expectation value of spins along the vector $a$.

\section{Discussions and Conclusions}

The essence of Bell's theorem is that the following problem of moments has
no solution (one takes $ab=\cos(\alpha-\beta)$ and $f^{(1)}=-f^{(2)}$ in
(\ref{eqn6x}))
\be
\label{mom}
\cos(\alpha-\beta)=\int f(\alpha, \omega) f(\beta, \omega) dP(\omega)
\ee
where one assumes
\be
\label{ineq}
|f(\alpha, \omega)\leq 1
\ee
Here the last condition (\ref{ineq}) is crucial. It means that one
describes spin by a single classical random variable which transforms as
{\it scalar} under rotations in space. However it is well known that spin
operator is not a scalar and it transforms as a vector under rotations.
Therefore if we want to describe spin by means of classical random
variables then it seems more natural to use not a single scalar random
variable as J.S. Bell did but to use a triple of random variables as  it
was done in the previous section.

If we relax the condition (\ref{ineq}) then one can solve the 
problem of moments. For example the following problem of moments
\be
\label{mom2}
\\cos(\alpha-\beta)=2
\int f(\alpha, \omega) f(\beta, \omega) dP(\omega)
\ee
has a solution
$$
\cos(\alpha-\beta)=2
\int_0^{2\pi}\cos(\alpha-\omega) \cos(\beta-\omega) \frac{d\omega}{2\pi}
$$
There is no contradiction between representation (\ref{eq00x}) and Bell's
theorem. Indeed if we take $f^{(1)}(a,\omega)= \xi_i^{(1)}(\omega)a_i$ and
$f^{(2)}(b,\omega)=\xi_i^{(2)}(\omega)b_i$ then the representation
(\ref{eq00x}) has the form of (\ref{eqn6x})  but now the condition
(\ref{ineq}) is not satisfied. 

The spectral theorem of von Neumann \cite{Yos} is relevant in this
discussion. It states that if $A_i$, $i=1,\ldots,n$ is a set of commuting
observables  (Hermitian operators in a Hilbert space $\EuScript{H}$) then
for any unit vector  $\psi \in \EuScript{H}$ there exists a representation
$$
\left<\psi|A_1\cdots A_n|\psi\right>
=\int f_1(\omega)\cdots f_n(\omega) dP(\omega)
$$
This representation is local in the sense that  every $f_i (\omega)$ is a
real function depending only on $A_i$ and on $\omega$. In particular one
can apply this theorem in the case when the condition of locality means
that $A_i$ is an operator  in a Hilbert space $\EuScript{H}_i$ and
$\EuScript{H}=\EuScript{H}_1\otimes\cdots\otimes\EuScript{H}_n$. 
Therefore  quantum correlation function of commutative observables can be
always represented as an expectation  value of local classical random
variables.

Let us summarize the main properties of the classical models of spin
discussed in this paper. In quantum mechanics spin is represented by three
Pauli matrices $\sigma_i$ with standard vector transformation rules under
rotations. In Bell's model spin is represented by the random field
$f(a,\omega)=\pm 1$. In this paper spin is represented by three random
variables $\xi_i(\omega)$ with appropriate transformation rules under
rotations. Bell's model of spin can not reproduce the quantum correlation
functions of two spins while the model with $\xi_i(\omega)$ can do it. 
 Although the model with $\xi_i(\omega)$ is classical
it does not provide a complete description of reality because
it refuses to describe results of individual measurements along
an arbitrary vector $a$. The model describes only results
of individual measurements along the three preferred orthogonal
vectors and also expectation values along an arbitrary vector. 

To conclude, in this paper the new classical model of spin is discussed
which perhaps can help in further considerations of problems of locality,
reality, and causality in quantum mechanics. In particular it would be
interesting to reconsider from this point of view the Einstein, Podolsky
and Rosen paradox which in its original form is not equivalent \cite{Pop}
to its Bohm's spin formulation used in Bell's consideration.

\section{Acknowledgments}

We are grateful to L.Accardi, I.Aref'eva, A.Baranov, Yu.Drozzinov,
A.Gusch\-chin and B.Zavialov for useful discussions on Bell's theorem.


\end{document}